\documentclass[10pt]{article}
\usepackage[dvips]{graphicx}

\usepackage{epsfig}
\usepackage{graphicx}
\usepackage{indentfirst}
\usepackage{amsmath}
\usepackage{amsfonts}
\usepackage{amssymb}

\begin{document}

\begin{center}
{\Large\bf  Reconstruction of  f(R, T) gravity  describing  matter dominated and accelerated phases\\}
\medskip

 M. J. S. Houndjo\footnote{e-mail:
sthoundjo@yahoo.fr}\\ 
 \,Instituto de F\'{i}sica, Universidade Federal da Bahia, 40210-340, Salvador, BA, Brazil

\end{center}

\begin{abstract}
We investigate the cosmological reconstruction in modified $f(R,T)$ gravity, where $R$ is the Ricci scalar and $T$ the trace of the stress-energy tensor. Special attention is attached to the case in which the function $f$ is given by $f(R, T)=f_1(R)+f_2(T)$. The use of auxiliary scalar field is considered with two known examples for the scale factor corresponding to an expanding universe. In the first example, where ordinary matter is usually neglected for obtaining the unification of matter dominated and accelerated phases with $f(R)$gravity, it is shown in this paper that this unification can be obtained without neglect ordinary matter. In the second example, as in $f(R)$gravity, model of $f(R, T)$ gravity with transition of matter dominated phase to the acceleration phase is obtained. In both cases, linear function of the trace is assumed for $f_2(T)$ and it is obtained  that  $f_1(R)$ is proportional to a power of $R$ with exponents depending on the input parameters.
 
\end{abstract}

Pacs numbers: 98.80.-k, 95.36.+x, 11.25.-w

\section{Introduction}
Recent observations data confirmed that the current expansion of the universe is accelerating \cite{auna, aunb, aunc, aund, adeuxa, adeuxb}. There are two approaches to explain this accelerated expansion of the universe. One is introducing some unknown element called dark energy in the framework of general relativity. In particular the five-year Wilkinson Microwave Anisotropy Probe (WMAP) data \cite{d1} give the bounds for the value of the equation of state (EoS) parameter $\omega_{DE}$, which is the ratio of the pressure of the dark energy to its energy density, in the range of $-1.11<\omega_{DE}<-0.86$. This could be consistent if the dark energy is a cosmological constant with $\omega_{DE}=-1$ and therefore our universe seems to approach asymptotically a de Sitter universe. There existed also a period of another accelerating expansion of the universe, called inflation, in the early universe. The second approach is to modify the gravitational theory, e.g., to study the action described by an arbitrary function of the scalar curvature $R$. This is called $f(R)$gravity, where $f(R)$ is an arbitrary function of the scalar curvature $R$ \cite{e1,e2a,e2b,e2c,e2d}. The present late-time cosmic acceleration of the universe  can be explained by $f(R)$ gravity \cite{a4}. Viable cosmological models have been found in \cite{a5a,a5b,a5c,a5d,a5e,a5f,a5g,a5h,a5i,a5j,a5k,a5l,a5m,a5n,a5o,a5p,a5q,a5r,a5s,a5t,a5u, a5v} under some conditions, and weak field constraints obtained from the classical tests of general relativity for the solar system regime seem to rule out most of the models so far \cite{a6a,a6b,a6c,a6d,a6e,a6f,a7}.  Other interesting class of modified gravity models which can easily reproduce the late-time acceleration epoch is string-inspired modified Gauss-Bonnet gravity, so called $f(G)$ gravity \cite{g33,g34a,g34b,g34c,g34e,g34f,g34g,g34h,g34i,g34j,g34k,g34l,g34m,g34n,g34o,g34p,g34q,g34r,g34s,g34t,g35,g36}, where $f(G)$ is an arbitrary function of the Gauss-Bonnet invariant $G= R^2-4R_{\mu\nu}R^{\mu\nu}+R_{\mu\nu\lambda\sigma}R^{\mu\nu\lambda\sigma}$ ($R_{\mu\nu}$ and $R_{\mu\nu\lambda\sigma}$ are the Ricci tensor and the Riemann tensor, respectively). In \cite{g}, $f(R,G)$ models has been used as modified gravity and in which accelerating cosmologies can realize the finite-time singularities. Recently, quantum effects from particle production around these finite-time singularities have been investigated in the framework of $f(R)$ \cite{vincentpaper}.\par 
However, there exists another extension of the standard general relativity, the $f(R,T)$ modified theories of gravity, where the Lagrangian is an arbitrary function of $R$ and $T$. The terminology used  here by denoting the trace of the energy momentum tensor by $T$ needs a special care and may not be confused with the current terminology, $f(T)$ theory \cite{daouda1,daouda2,rafael,baodjiu,christian,cemsinan}, in which $T$ denotes the torsion scalar.\par

 The dependence from $T$ (the trace) may be induced by exotic imperfect fluids or quantum effects (conformal anomaly). This modified gravity has been used recently in \cite{a}. They obtained the gravitational field equations in the metric formalism, as well as the equations of motion for test particles, which follow from the covariant divergence of the stress-energy tensor. The field equations of several particular models, corresponding to some explicit forms of the function $f(R, T)$ have been presented. Therefore, they obtained that the predictions of the $f(R, T)$ gravity model could lead to some major differences in several problems of current interest, such as  cosmology, gravitational collapse or the generation of gravitational waves. The study of these phenomena may also provide some specific signatures and effects, which could distinguish and discriminate between various gravitational models.\par

One of the interesting point in cosmology is the reconstruction program for modified gravity. The known classical universe expansion history can be used for the explicit and successful reconstruction of some versions (of some form or with specific potentials) from $f(R)$, $f(G)$ and $f(R, G)$ gravities \cite{g,c,b}. In \cite{f}, it is demonstrated that cosmological sequence of matter dominance, deceleration-acceleration transition and acceleration era may always emerge as cosmological solutions of such theory. This reconstruction technique has been used in \cite{b} and it is demonstrated in $f(R)$ gravity framework that there are models where any known (big rip, II, III, or IV type) singularity may classically occur. In \cite{c}, also in $f(R)$ framework, several versions of modified gravity compatible with Solar System tests are presented  with the occurrence of the above sequence of cosmological epoch. It is also shown in alternative approach that simple version of $f(R)$ modified gravity may lead to the unification of matter dominated and accelerated phases. \par

In this paper, we propose to develop the general  reconstruction program for $f(R, T)$ modified gravity. We use the representation with the auxiliary scalar field.  Two known examples for the scale factor corresponding to an expanding universe are considered. With the first example, we find the possibility of unification of matter-dominated and accelerated phases with $f(R, T)$ modified gravity. Remember that in $f(R)$ gravity, with this example for the scale factor, ordinary matter is usually neglected for simplicity and $f(R)$ terms contribution plays the role of matter. In this case, contrary to $f(R)$ modified gravity, it is not necessary to neglect ordinary matter contribution and naturally the unification scheme with $f(R, T)$ gravity is obtained. In the second example, we find the existence of $f(R, T)$ gravity models with transition of matter dominated phase to the acceleration phase.  In both cases, linear function of the trace is assumed for $f_2(T)$ and it is obtained  that  $f_1(R)$ is proportional to a power of $R$ with exponents depending on the input parameters..\par
The paper is organized as follows. In the second section, we present the general formulation of the reconstruction scheme of $f(R, T)$ modified gravity. In the third section, using a first example of the scale factor corresponding to an expanding universe, we present the possibility of unification of matter dominated and accelerated phases with $f(R, T)$ gravity. A model of $f(R, T)$ gravity with transition of matter dominated to the accelerated phase is presented in the fourth section and the conclusion and perspective in the fifth section.

\section{General formulation}

Let us consider the general reconstruction scheme for modified gravity with  f(R, T) action
\begin{eqnarray}\label{e1}
S=\int d^4x\sqrt{-g}\lbrace f(R, T)+\mathcal{L}_{m}\rbrace\,\,,
\end{eqnarray}
where $f(R, T)$ is an arbitrary function of the scalar curvature $R$ and $T$ the trace of the energy momentum tensor, defined from the matter Lagrangian density $\mathcal{L}_m$ by
\begin{eqnarray}\label{e2}
T_{\mu\nu}=-\frac{2}{\sqrt{-g}}\frac{\delta\left( \sqrt{-g}\mathcal{L}_m\right) }{\delta g^{\mu\nu}}\,\,\,.
\end{eqnarray}
First we consider the proper Hubble parameter $H$, which describes the evolution of the universe with matter dominated era and accelerating expansion. One can find $f(R, T)$ theory
realizing such a cosmology. \par
One  shows that it is possible to obtain any cosmology reconstructing a function  $f(R, T)$ on theoretical level.\par 
The equivalent form of above action  can be written as
\begin{eqnarray}\label{e3}
S=\int d^4x\sqrt{-g}\lbrace P_1(\phi)R+P_2(\phi)T+Q(\phi)+\mathcal{L}_m\rbrace\,\,,
\end{eqnarray}
where $P_1$, $P_2$ and Q are proper functions of the scalar field $\phi$ and $\mathcal{L}_m$ is the matter Lagrangian density. We assume that  the scalar field $\phi$ does not have a kinetic term and may be regarded as an auxiliary field. Then, by the variation other  $\phi$ , one obtains
\begin{eqnarray}\label{e4}
P_1^{\,\prime}(\phi)R+P_2^{\,\prime}(\phi)T+Q^{\,\prime}(\phi)=0\,,
\end{eqnarray}
which may be solved with respect to $\phi$:
\begin{eqnarray}\label{e5}
\phi=\phi(R, T)\,\,.
\end{eqnarray}
Substituting  (\ref{e5}) into (\ref{e3}), one obtains $f(R, T)$-gravity :
\begin{eqnarray}\label{e6}
S=\int d^4x\sqrt{-g}\lbrace f(R, T) +\mathcal{L}_m\rbrace\,,\quad f(R, T)\equiv P_1[\phi(R, T)]R+P_2[\phi(R, T)]T+Q[\phi(R, T)]
\end{eqnarray}
By varying the action $S$ with respect to the metric one obtains \cite{a},
\begin{eqnarray}\label{e7}
f_R(R, T)R_{\mu\nu}-\frac{1}{2}f(R, T)g_{\mu\nu}+\left( g_{\mu\nu}\Box -\nabla_\mu\nabla_\nu\right) f_R(R, T)=\frac{1}{2} T_{\mu\nu}-f_T(R, T)T_{\mu\nu}-f_T(R, T)\Theta_{\mu\nu}\,\,,
\end{eqnarray}
where $\Theta_{\mu\nu}$ is defined by 
\begin{eqnarray}\label{e8}
\Theta_{\mu\nu}\equiv g^{\alpha\beta}\frac{\delta T_{\alpha\beta}}{\delta g^{\mu\nu}}= -2T_{\mu\nu}+g_{\mu\nu}\mathcal{L}_m-2g^{\alpha\beta}\frac{\partial^{2}\mathcal{L}_m}{\partial g^{\mu\nu}\partial ^{\alpha\beta}}\,\,.
\end{eqnarray}
Here $f_R$ and $f_T$ denote the derivatives of  $f$ with respect to $R$ and $T$, respectively. 
Let us assume for simplicity that the function $f$ is given by $f(R, T)=f_1(R)+f_2(T)$, where $f_1(R)$ and $f_2(T)$ are arbitrary functions of $R$ and $T$, respectively. Then, one can re-write (\ref{e7}) as 
\begin{eqnarray}\label{e9}
f_{1R}(R)R_{\mu\nu}-\frac{1}{2}f_1(R)g_{\mu\nu}+\left( g_{\mu\nu}\Box-\nabla_\mu\nabla_\nu\right) f_{1R}(R)=\frac{1}{2}T_{\mu\nu}-f_{2T}(T)T_{\mu\nu}-f_{2T}(T)\Theta_{\mu\nu}+\frac{1}{2}f_2(T)g_{\mu\nu}\,\,.
\end{eqnarray}
Assuming that the matter content is a perfect fluid, the stress tensor is given by
\begin{eqnarray}\label{e10}
T_{\mu\nu}=(\rho+p)u_\mu u_\nu-pg_{\mu\nu}\,\,,
\end{eqnarray}
where $u_{\mu}$ is the four-velocity which satisfies the condition $u_\mu u^{\mu}=1$. Then,  the matter Lagrangian density can be taken as $\mathcal{L}_m=-p$, and  $\Theta_{\mu\nu}=-2T_{\mu\nu}-pg_{\mu\nu}$. Thus,  the equation (\ref{e9}) becomes
\begin{eqnarray}\label{e11}
f_{1R}(R)R_{\mu\nu}-\frac{1}{2}f_1(R)g_{\mu\nu}+\left( g_{\mu\nu}\Box-\nabla_\mu\nabla_\nu\right) f_{1R}(R)=\frac{1}{2}T_{\mu\nu}+f_{2T}(T)T_{\mu\nu}+\left[ f_{2T}(T)p+\frac{1}{2}f_2(T)\right] g_{\mu\nu}\,\,.
\end{eqnarray}
Now, choosing the functions $f_1$ and $f_2$ as
\begin{eqnarray}\label{e12}
f_1(R)=P_1(\phi)R+Q_1(\phi)\,\,,\quad f_2(T)=P_2(\phi)T+Q_2(\phi)\,\,, \quad  Q_1(\phi)+Q_2(\phi)=Q(\phi)\,\,,
\end{eqnarray}
one can write the time-time and space-space components of the field equation (\ref{e11}) as
\begin{eqnarray}
6P_1(\phi)H^2-Q(\phi)+6H\dot{P}_1(\phi)=\rho+P_2(\phi)\left( 3\rho-p\right)\label{e13} \,\,,\\
-4H\dot{P}_1(\phi)-2\ddot{P}_1(\phi)-4\dot{H}P_1(\phi)-6H^2P_1(\phi)+Q(\phi)=p-P_2(\phi)\left( \rho-3p\right) \label{e14}\,\,,
\end{eqnarray}
where the dot is the derivative with respect to the cosmic time $t$.\par
We have to determine the respective expressions of $P_1(\phi)$, $P_2(\phi)$ and $Q(\phi)$. Thus, by combining the equations (\ref{e13}) and (\ref{e14}) and cancelling $Q(\phi)$, we obtain the following equation
\begin{eqnarray}\label{e15}
2H\dot{P}_1(\phi)-2\ddot{P}_1-4\dot{H}P_1(\phi)=\left( \rho+p\right) \left[ 1+2P_2(\phi)\right] \,\,.
\end{eqnarray}
Using (\ref{e13}) and (\ref{e14}), we can determine the effective energy density and pressure as
\begin{eqnarray}
\rho_{eff}=-6H^2(P_1(\phi)-1)+Q(\phi)-6H\dot{P}_1(\phi)+\rho+(3\rho-p)P_2(\phi)\,\,,\label{e16}\\
p_{eff}=6H^2(P_1(\phi)-1)+4\dot{H}(P_1(\phi)-1)-Q(\phi)+2\ddot{P}_1(\phi)+4H\dot{P}_1(\phi)+p-(\rho-3p)P_2(\phi)\,\,.\label{e17}
\end{eqnarray} 
Note that  the effective energy density and pressure are functions of the ordinary energy density $p$ and ordinary pressure $p$. This is a effect of the presence of the trace if the gravitational part of the action. For general relativity with $f(R, T)= R$, $\rho_{eff}=\rho$ and $p_{eff}=p$ and therefore Eqs. (\ref{e16}) and (\ref{e17}) are Friedmann equations. Consequently, Eqs. (\ref{e16}) and (\ref{e17}) imply that the contribution of modified gravity can formally be included in the effective energy and pressure of the universe.  Assuming that the effective energy density and pressure, and ordinary energy density and pressure satisfy the conservation law separately, one obtains
\begin{eqnarray}
-12H^2\dot{P}_1(\phi)+\dot{Q}(\phi)-6\dot{H}\dot{P}_1(\phi)+3H\rho(\omega^2-1)P_2(\phi)+\rho(3-\omega)\dot{P}_2(\phi)=0\,\,,\label{e18}
\end{eqnarray}
where we used $p=\omega\rho$. Deriving the equation (\ref{e16}) and combining it with (\ref{e18}), the equation (\ref{e15}) is recovered. \par
As one can redefine the scalar field properly, we may choose it as
\begin{eqnarray}\label{e19}
\phi=t\,\,.
\end{eqnarray}
If the scale factor is given by a proper function g(t) as
\begin{eqnarray}\label{e20}
a(t)=a_0e^{g(t)}\,\,,
\end{eqnarray}
the equation (\ref{e15}) can be written as 
\begin{eqnarray}\label{e21}
-4\ddot{g}(\phi)P_1(\phi)+2\dot{g}(\phi)\frac{d {P}_1(\phi)}{d\phi}-2\frac{d^2{P}_1(\phi)}{d\phi^2}=\rho(1+\omega)(1+2P_2(\phi))\,\,.
\end{eqnarray}
Fixing a value for $P_2(\phi)$ and  solving  (\ref{e21}), we can find the form of $P_1(\phi)$. Using (\ref{e13}) one can determine $Q(\phi)$ as 
\begin{eqnarray}
Q(\phi)= 6H^2P_1(\phi)+6H\dot{P}_1-\rho-\rho(3-\omega)P_2(\phi)\,\,.\label{e22}
\end{eqnarray}
Thus, in principle, any cosmology expressed as (\ref{e20}) can be  formulated by some specific f(R, T)-gravity. 
\section{Unification of matter dominated and accelerated phases with f(R, T) gravity }
In this section we fix $P_2(\phi)=-\frac{1}{2}$ and consider the following example,
\begin{eqnarray}\label{e23}
\dot{g}(\phi)=g_0+\frac{g_1}{\phi}\,\,,
\end{eqnarray}
with which , Eq.(\ref{e21}) becomes 
\begin{eqnarray}\label{e24}
2\frac{g_1}{\phi^2}P_1+\left( g_0+\frac{g_1}{\phi}\right) \frac{dP_1}{d\phi}-\frac{d^2P_1}{d\phi^2}=0\,\,.
\end{eqnarray}
Putting $P_1(\phi)=\phi^q\lambda(\phi)$, one obtains the following equation
\begin{eqnarray}\label{e25}
\phi\ddot{\lambda}+ \left(2q-g_1-g_0\phi\right) \dot{\lambda}-\left[ \frac{g_1(2+q)-q(q-1)}{\phi}+g_0q\right] \lambda=0
\end{eqnarray}
which leads to a confluent hyper-geometric functions as solutions when $g_1(2+q)-q(q-1)=0$. Then, one has as solutions of (\ref{e24}),
\begin{eqnarray}\label{e26}
P_{1,1}(\phi)=\phi^q\,_1F_1(q, 2q-g_1, g_0\phi)\,\,,\quad P_{1,2}=\phi^{1-q+g_1}\,_1F_1(1-q+g_1, 2-2q+g_1; g_0\phi)\,\,,
\end{eqnarray}
where 
\begin{eqnarray}\label{e27}
q=\frac{1}{2}\left[ 1+g_1\pm\sqrt{1+10g_1+g_1^2}\right] \,\,.
\end{eqnarray}
From the Eq.(\ref{e23}) one sees that the Hubble parameter is given by
\begin{eqnarray}\label{e28}
H=g_0+\frac{g_1}{t}\,\,.
\end{eqnarray}
It follows that when $t$ is small,  $H\sim g_1/t$, the  universe is filled with a perfect fluid with the EoS parameter $\omega=-1+2/(3g_1)$. However, when the $t$ is large enough, the Hubble parameter tends to a constant ($H\rightarrow g_0$) and the universe looks as deSitter one. This is the real possibility of the transition from the matter dominated phase to the accelerating one. Note that in the same way, one can construct $f(R, T)$ action describing other epoch remembering that form of modified gravity is different for different epochs (  the inflationary epoch action is different from the form at late-time universe).\par 
We can now investigate the asymptotic forms of $f(R, T)$ in (\ref{e6}) corresponding to  (\ref{e23}). When $\phi$ (or t) is small, we find
\begin{eqnarray}\label{e29}
P_1\sim P_0\phi^q\,\,,   \quad P_0=\mbox{const}\,\,.
\end{eqnarray}
Using (\ref{e22}), one finds 
\begin{eqnarray}\label{e30}
 Q\sim\frac{1}{2}(1-\omega)\rho_0a_0^{-3(1+\omega)}\phi^{-3(1+\omega)g_1}+6P_0g_1(g_1+q)\phi^{q-2}\,\,,
\end{eqnarray}
where we used $a=a_0e^{g_0\phi}\phi^{g_1}$ and $\rho=\rho_0a^{-3(1+\omega)}$. Depending on the value of the parameter $\omega$  one can distinguish two cases. \par 
$\bullet$ If $q<2-3(1+\omega)g_1$, one has
\begin{eqnarray}\label{e31}
Q\sim  6P_0g_1(g_1+q)\phi^{q-2}\,\,,
\end{eqnarray}
and using (\ref{e4}), it follows that
\begin{eqnarray}\label{e32}
\phi^2 \sim \frac{6g_1(g_1+q)(2-q)}{qR}\,\,\,.
\end{eqnarray}
This result is very similar to that obtained in $ f(R)$ gravity \cite{c}, where the first term in the right side of (\ref{e30}) for Q does not appear and the input parameters are not the same as in this case. Moreover, it is important to note that the result in \cite{c} is found with vanishing the energy density and the pressure of the ordinary matter, considering that the $f(R)$ terms contribution play the role of matter. 
In the $ f(R, T)$ gravity, vanishing the ordinary energy density and pressure is not necessary for reconstructing a viable action. This shows that some cases of $f(R, T)$-gravity are more general, in which any restriction on the content of the universe is necessary.  
With (\ref{e32}), one has 
\begin{eqnarray}\label{e33}
P_1(\phi)R\sim P_0\left[ \frac{6g_1(g_1+q)(2-q)}{q}\right] ^{\frac{q}{2}}R^{-\frac{q}{2}+1}\,\,,\quad Q(\phi)\sim  6P_0g_1(g_1+q)\left[ \frac{6g_1(g_1+q)(2-q)}{q}\right]^{\frac{q}{2}-1}R^{-\frac{q}{2}+1}\,\,.
\end{eqnarray}
Combining the expressions of  (\ref{e31}), one obtains
\begin{eqnarray}\label{e34}
f(R, T) \sim \frac{2P_0}{q-2}\left[ \frac{6g_1(g_1+q)(q-2)}{q}\right] ^{\frac{q}{2}}R^{-\frac{q}{2}+1}-\frac{1}{2}T \,\,.
\end{eqnarray}
$\bullet$ If $q>2-3(1+\omega)g_1$ one has
\begin{eqnarray}\label{e35}
Q\sim \frac{1}{2}(1-\omega)\rho_0a_0^{-3(1+\omega)}\phi^{-3(1+\omega)g_1}
\end{eqnarray}
and with the use of (\ref{e4}) we obtain
\begin{eqnarray}\label{e36}
\phi \sim\left[ \frac{3(1-\omega^2)g_1\rho_0a_0^{-3(1+\omega)}}{2qP_0}\right] ^{\frac{1}{q+3(1+\omega)g_1}}R^{-\frac{1}{q+3(1+\omega)g_1}}\,\,\,.
\end{eqnarray}
Then, one gets
\begin{eqnarray}\label{e37}
P(\phi)R\sim P_0\left[ \frac{3(1-\omega^2)g_1\rho_0a_0^{-3(1+\omega)}}{2qP_0}\right] ^{\frac{q}{q+3(1+\omega)g_1}}R^{\frac{3(1+\omega)g_1}{q+3(1+\omega)g_1}}\,\,\,,\nonumber\\
Q(\phi)\sim \frac{1}{2}(1-\omega)\rho_0a_0^{-3(1+\omega)}\left[ \frac{3(1-\omega^2)g_1\rho_0a_0^{-3(1+\omega)}}{2qP_0}\right] ^{\frac{-3(1+\omega)}{q+3(1+\omega)g_1}}R^{\frac{3(1+\omega)g_1}{q+3(1+\omega)g_1}}\,\,\,.
\end{eqnarray}
Combining the expressions of (\ref{e37}), one gets
\begin{eqnarray}\label{e38}
f(R, T)\sim \frac{P_0\left[ q+2(1+\omega)g_1\right] }{2(1+\omega)g_1}\left[ \frac{3(1-\omega^2)g_1\rho_0a_0^{-3(1+\omega)}}{2qP_0}\right] ^{\frac{q}{q+3(1+\omega)g_1}}R^{\frac{3(1+\omega)g_1}{q+3(1+\omega)g_1}}-\frac{1}{2}T\,\,\,.
\end{eqnarray}
On the other hand, when $g_0\phi$ (or t) is large and positive, one gets
\begin{eqnarray}
P_1&\sim& \tilde{P}_0\phi^{g_1}e^{g_0\phi}\left[1+\frac{(q-g_1)(1-q)}{g_0 \phi}\right]\,\,\,, \label{e39}\\
Q&\sim& 12\tilde{P}_0 g_0^2e^{g_0\phi} \phi^{g_1}\left[ 1+\frac{g_1+q+g_1q-q^2}{g_0\phi}\right] \,\,.\label{e40}
\end{eqnarray}
Using (\ref{e4}) and the expressions  (\ref{e39}) and (\ref{e40}), one obtains
\begin{eqnarray}\label{e41}
\phi &\sim& \frac{K}{\frac{R}{12g_0^2}+1}\,\,,\quad K= q^2-g_1q-q-2g_1\,\,.
\end{eqnarray}
Then we find 
\begin{eqnarray}\label{e42}
f(R, T)\sim 12\tilde{P}_0 g_0 ^2K^{g_1} \left(1+\frac{R}{12g_0^2}\right)^{1-g_1}\exp{\left( \frac{K}{\frac{R}{12g_0^2}+1}\right)}-\frac{1}{2}T\,\,\,.
\end{eqnarray} 
This shows that using  a specific given model of  f(R, T) gravity, there is a strong possibility of unification of matter-dominated phase, transition to acceleration and late time speed up of the universe.
\section{Model of $f(R, T)$ gravity with transition of matter dominated to the accelerated phase}
 Let us consider a realistic example where the total action contains also the usual matter. The function $g(\phi)$ is putting into the form 
 \begin{eqnarray}\label{e43}
 g(\phi)=h(\phi)\ln{( \phi)}\,\,\,.
 \end{eqnarray} 
 We will assume that $h(\phi)$ is a slowly changing function of $\phi$. Using the adiabatic approximation, we can neglect the derivatives of $h(\phi)\, ( h^\prime(\phi) \sim h^{\prime\prime}\sim 0)$.  Redefining $P_1(\phi)$ as  $P_1(\phi)=e^{g(\phi)/2}s(\phi)$ and fixing $P_2(\phi)=-\frac{1}{2}$ , Eq. (\ref{e21})  becomes
\begin{eqnarray}\label{e44}
s^{\prime\prime}-\left[ \frac{h^2(\phi)+10h(\phi)}{4\phi^2}\right] s=0\,\,.
\end{eqnarray}
The general solution of (\ref{e44}) is written as 
\begin{eqnarray}\label{e45}
s(\phi)=s_{+}\phi^{\sigma_+(\phi)}+s_-\phi^{\sigma_-(\phi)}\,\,\,,
\end{eqnarray}
where $s_{\pm}$ are arbitrary constants  and  $\sigma_{\pm}$ are defined by
\begin{eqnarray}\label{e46}
\sigma_{\pm}(\phi)=\frac{1}{2}\left[ 1\pm \sqrt{1+h^2(\phi)+10h(\phi)}\right] \,\,\,.
\end{eqnarray}
Thus, $P_1(\phi)$ is
\begin{eqnarray}\label{e47}
P_1(\phi)=P_{1+}\phi^{\Sigma_+(\phi)}+P_{1-}\phi^{\Sigma_-(\phi)}\,\,\,,
\end{eqnarray}
where $P_{1\pm}$ are arbitrary constants proportional to $s_{\pm}$ and $\Sigma_{\pm}$ defined by
\begin{eqnarray}\label{e48}
\Sigma_ {\pm}=\frac{h(\phi)}{2}+\sigma_{\pm}\,\,\,\,.
\end{eqnarray}
From (\ref{e43}), we get for the Hubble parameter
 and the scalar curvature respectively
\begin{eqnarray}\label{e49}
H\sim\frac{h(t)}{t}\,\,, \quad R\sim \frac{6\left(- 2h(t)^2+h(t)\right) }{t^2}\,\,\,\,.
\end{eqnarray} 
Let us consider the following form for the function $h(\phi)$,
\begin{eqnarray}\label{e50}
h(\phi)=\frac{h_i+h_fq\phi^2}{1+q\phi^2}\,\,\,,
\end{eqnarray}
where $q$, $h_i$ and $h_f$ are constants. Note that when $\phi\rightarrow 0$, $h(\phi)\rightarrow h_i$ while for $\phi\rightarrow \infty$, $h(\phi)\rightarrow h_f$. Also, its easy to show that $\ddot{a}/a =h(\phi)(-1+h(\phi))/\phi^2$. Then, if $0<h_i<1$, the early universe is in deceleration phase and if  $h_f$, the late universe is in an acceleration phase. Consider  the situation where $h(\phi)\sim h_m$ is almost constant when $\phi\sim t$, with $0<< t_m<<+\infty$. If $h_i>1$, $h_f>1$ and $0<h_m<1$, the early universe is an accelerating phase which would be  inflation. After that, the universe enters in a decelerating phase, which correspond to matter dominated phase with $h(\phi)\sim 2/3$ there. Later, the universe enter in accelerating phase, which would be the late-time acceleration. Remark that when $q$ is small enough, $h(\phi)$ can be slowly varying function of $\phi$.\par 
Using Eqs.(\ref{e49}) and (\ref{e50}), one obtains three solutions for $\phi^2$,
\begin{eqnarray}
\phi^2_0=\frac{1}{6q^2 R}\left[ J(R)+K(R)\left[ L(R)+M(R)\right] ^{-1/3}+2\left[ L(R)+M(R)\right] ^{1/3}\right] \,\,,\label{e51}\\
\phi^2_{\pm}=\frac{1}{6q^2R}\left[ J(R)+e^{\mp 2i\pi/3}K(R)\left[ L(R)+M(R)\right] ^{-1/3}+2e^{\pm 2i\pi/3}\left[ L(R)+M(R)\right] ^{1/3}\right]\,\,,\label{e52}
\end{eqnarray}
where
\begin{eqnarray}
J(R)=4q(-3h_fq+6h_f^2q-R)\,\,,\nonumber
\end{eqnarray}
\begin{eqnarray}
K(R)=2q^2\left[-144h_f^3q^2+144h_f^4q^2+12h_f^2q(3 q-4R)+6h_f(1+12h_i)qR+R(-18h_iq+R)\right]\,\,,\nonumber
\end{eqnarray}
\begin{eqnarray}
L(R)=\Bigg[-2592h_f^5q^6+1728h_f^6q^6+432h_f^4q^5(3 q-2R)-18h_f^2q^4\left[(3+54h_i)q-5R\right]R \nonumber\\- 27h_iq^4R^2+162h_i^2q^4R^2+q^3R^3+9h_fq^4R\left[6h_i(3q -4R)+R\right]-108h_f^3q^5\left[2q-(5+12h_i)R\right]\Bigg]\,\,,\nonumber
\end{eqnarray}
\begin{eqnarray}
M(R)=\Bigg[-(h_f-h_i)^2q^7R^2\Bigg(432h_f^4q^3-144h_f^3q^2\left[3q +(-1+4h_i)R\right]+12h_fqR\left[3(1+6h_i)q+(-5+18h_i)R\right]\nonumber\\
+R\left[-36h_iq(2q-3R)-324h_i^2qR+(3q-4R)R\right]+ 12h_f^2q\left[9q^2+12(-2+3h_i)qR+R^2\right]\Bigg)\Bigg]^{1/9} \,\,.\nonumber
\end{eqnarray} 
From Eq.(\ref{e49}) and (\ref{e50}) it follows that the $\phi$ (or t) is large, the curvature $R\sim 6(-2h_f^2+h_f)/\phi^2$ is small and  when $\phi$ (or t) is small $R\sim 6(-2h_i^2+h_i)/\phi^2$ is large. Thus,  one can investigate the asymptotic forms of $f(R, T)$ corresponding to (\ref{e43}). Note that, always, $\Sigma_+> \Sigma_-$.\par 
Then, when $\phi$ is small, the suitable expression for $P_1$ is 
\begin{eqnarray}\label{e53}
P_1(\phi)=P_{1-}\phi^{\Sigma_-(\phi)}\,\,.
\end{eqnarray}
Assuming a value of $h_i$ such that $\Sigma_->2-3(1+\omega)h_i$ and using (\ref{e22}), one obtains
\begin{eqnarray}\label{e54}
Q(\phi)\sim\frac{1}{2}(1-\omega)\rho_0\phi^{-3(1+\omega)h_i}\,\,,
\end{eqnarray}
and using (\ref{e4}) one gets the expression
\begin{eqnarray}\label{e55}
f(R, T)\sim \frac{P_{1-}\left[\Sigma_-+3(1+\omega)h_i\right]}{3(1+\omega)h_i}\left[\frac{3\rho_0(1-\omega^2)h_i}{2P_{1-}\Sigma_-}\right]^{\frac{\Sigma_-}{\Sigma_-+3(1+\omega)h_i}}R^{\frac{3(1+\omega)h_i}{\Sigma_-+3(1+\omega)h_i}}-\frac{1}{2}T\,\,\,.
\end{eqnarray}
When $\phi$ is large, the dominant  term in (\ref{e47}) is 
\begin{eqnarray}\label{e56}
P_1(\phi)=P_{1+}\phi^{\Sigma_+ (\phi)}\,\,.
\end{eqnarray}
Assuming a value of $h_f$ such that $\Sigma_+< 2-3(1+\omega)h_f$,  one gets from (\ref{e22}),
\begin{eqnarray}\label{e57}
Q(\phi)\sim\frac{1}{2}(1-\omega)\rho_0\phi^{-3(1+\omega)h_f}\,\,\,,
\end{eqnarray}
and using (\ref{e4}), the corresponding gravitational expression is
\begin{eqnarray}\label{e58}
f(R, T)\sim \frac{P_{1+}\left[\Sigma_++3(1+\omega)h_f\right]}{3(1+\omega)h_f}\left[\frac{3\rho_0(1-\omega^2)h_f}{2P_{1+}\Sigma_+}\right]^{\frac{\Sigma_+}{\Sigma_++3(1+\omega)h_f}}R^{\frac{3(1+\omega)h_f}{\Sigma_++3(1+\omega)h_f}}-\frac{1}{2}T\,\,\,\end{eqnarray}
If the early universe is a matter dominated phase, then $h_i=2/3$ and the expression (\ref{e55}) becomes

\begin{eqnarray}\label{e59}
f(R, T)\sim \frac{P_{1-}(17-\sqrt{73}+12\omega)}{12(1+\omega)} \left[\frac{6\rho_0(1-\omega^2)}{P_{1-}(5-\sqrt{73})}\right]^{\frac{5-\sqrt{73}}{17-\sqrt{73}+12\omega}} R^{\frac{12(1+\omega)}{17-\sqrt{73}+12\omega}}-\frac{1}{2}T
\end{eqnarray}
With this model one can identify $\omega_{DE}=-1+2/(3h_f)$. As we mentioned in the introduction, from five year WMAP data, $-1.11<\omega_{DE}<-0.86$. Then, for $h_f$ large, $\omega_{DE}$ belongs to the above interval.
Thus, in this model, the early universe is a  matter dominated phase with f(R, T) as (\ref{e59}), which evolves into acceleration phase at late time with f(R, T) gravity corresponding to (\ref{e58}), and this is consistent with five years WMAP.\par
Despite the reconstruction results we obtained for $f(R,T)$, it is important to note that there exists a subclass of this theory which is not cosmological viable. The example that we used for $f(R,T)$ in this paper lets the comprehension more simple, that is $f(R,T)=f_1(R)+f_2(T)$. Observe that when $f_1(R)$ is linear in $R$, the left side of Eq.(\ref{e9}) is the same as in General Relativity. In such a situation, when the the stress tensor $T_{\mu\nu}$ is that of a field, the equation of motion (\ref{e9}) contains second order derivatives of the metric $g_{\mu\nu}$ and third order derivatives of matter field. This situation has already been studied in Palatini $f(R)$ gravity, where the presence of the higher order derivatives, leading to the appearance of singularities at discontinuities or irregularities in the matter distribution \cite{e2b,thomas}. It appears clearly that this risk of the appearance of singularities is avoided when $f_1(R)$ is non-linear in $R$.

\section{Conclusion}
We develop the cosmological reconstruction method for f(R, T) modified gravity for any given FRW metric. The resulting action is given in terms of special functions. The function f(R, T) is given as a sum of two arbitrary functions of R and T, $f_1(R)$ and $f_2(T)$ respectively. Two known examples corresponding to an expanding universe are used and the asymptotic forms of $f(R, T)$ investigated. It is obtained specific $f(R, T)$ gravity which describes the sequence of cosmological epochs: matter dominated stage and
accelerated one. In the first example, we presented the possibility of unification of matter dominated and accelerated  phases. In this case, an interesting point is that, contrary to f(R) gravity where ordinary matter is usually neglected, f(R, T) gravity reproduce this unification without any restriction. In the second example, it is shown that f(R, T) gravity can reproduce the transition of matter dominated phase to the accelerated one and this is consistent with five years WMAP data. In both cases, linear function of the trace is assumed for $f_2(T)$ and it is obtained  that  $f_1(R)$ is proportional to a power of $R$ with exponents depending on the input parameters. Despite these results, this theory needs particular attention because there exists a subclass for which the model is non-viable. Thus, for avoiding the appearance of singularities the function $f_1(R)$ may be non-linear in $R$.  \par
However, several investigations remain to be done in this theory. It would be interesting to discuss model which can reproduce $\Lambda CDM$-type cosmology, the stability and instability of the cosmological solutions and the corrections to Newton law as usually done in f(R) and f(G) gravities.

\vspace{1cm}

{\bf Acknowledgement}:   The author thanks  CNPq-Brazil for partial financial support.

\end{document}